# WRIST BONE SEGMENTATION IN X-RAY IMAGES USING CT-BASED SIMULATIONS

*Youssef ElTantawy[1], Alexia Karantana[2], Xin Chen[1]*

[1]School of Computer Science, University of Nottingham, UK
[2]School of Medicine, University of Nottingham, UK

## ABSTRACT

Plain X-ray is one of the most common image modalities for clinical diagnosis (e.g. bone fracture, pneumonia, cancer screening, etc.). X-ray image segmentation is an essential step for many computer-aided diagnostic systems, yet it remains challenging. Deep-learning-based methods have achieved superior performance in medical image segmentation tasks but often require a large amount of high-quality annotated data for model training. Providing such an annotated dataset is not only time-consuming but also requires a high level of expertise. This is particularly challenging in wrist bone segmentation in X-rays, due to the interposition of multiple small carpal bones in the image. To overcome the data annotation issue, this work utilizes a large number of simulated X-ray images generated from Computed Tomography (CT) volumes with their corresponding 10 bone labels to train a deep learning-based model for wrist bone segmentation in real X-ray images. The proposed method was evaluated using both simulated images and real images. The method achieved Dice scores ranging from 0.80 to 0.92 for the simulated dataset generated from different view angles. Qualitative analysis of the segmentation results of the real X-ray images also demonstrated the superior performance of the trained model. The trained model and X-ray simulation code are freely available for research purposes: the link will be provided upon acceptance.

## 1. INTRODUCTION

Plain X-ray image (also known as 2D radiograph) is utilized in a wide variety of medical applications, such as bone fracture detection, pneumonia diagnosis, and cancer screening, due to its relatively low cost, quick image acquisition time, and capability to see-through the body. However, imaging the body using X-rays is complicated and sometimes ambiguous to be interpreted due to that 3D anatomic structures overlap each other in the 2D image. Hence, understanding the underlying 3D structures and detecting abnormalities from these 2D projections are challenging.

In computer-aided diagnosis, medical image segmentation is a fundamental step that helps identify the region of interest (e.g. organ, tumor, etc.). Due to the issues mentioned earlier, the automatic segmentation of X-ray images is challenging particularly when multiple anatomical structures are involved, such as the wrist joint (i.e. 8 carpal bones, 5 metacarpal bones, and distal radius and ulna).

Most existing works for wrist bone segmentation were based on conventional image processing methods, such as region growing [1], K-means variational level set [2], and statistical shape model [3]. They cannot work robustly on all the wrist bones and different view angles. Recently, machine learning methods, particularly deep learning models, have achieved state-of-the-art (SOTA) performance for automatic medical image segmentation. However, these methods require a large number of examples, including paired images and their corresponding segmentation masks, to train models. For a plain radiograph containing complicated structures like the wrist bones, manual labeling to generate the segmentation mask is time-consuming and requires a high level of expertise. The annotation accuracy cannot be guaranteed. The labeling task becomes much more challenging when the view of the wrist changes from an anterior-posterior (AP) view to a lateral view when multiple bones are overlapped in a small region. Despite these challenges, some existing works to train deep learning models used manually labeled datasets, as this is the most reliable and straightforward process. Kang et al [4] proposed a method for segmenting ten wrist bones utilizing a convolutional neural network (CNN). The model was trained and tested on locally collected 702 wrist X-ray images that were manually labeled. Notably, these images were all from AP view rather than the more challenging lateral views and their annotated dataset is not publicly available. Lee et al. [5] utilized the segment anything model (SAM) to annotate metacarpal bones using manually added prompts. This is a semi-automatic process and they excluded the eight carpal bones. In summary, no existing method can automatically segment all eight carpal bones and the two distal forearm bones in plain X-rays with flexible view angles. Such a method will provide tremendous help in wrist fracture detection, bone age assessment, and rheumatoid arthritis diagnosis.

In this paper, we propose a novel solution to train a deep-learning segmentation model to achieve automatic wrist bone segmentation in plain radiographs, which includes the following contributions. (1) The training images and their corresponding segmentation masks were simulated from wrist CT scans rather than using real X-ray images. It can provide a large amount of X-ray images and masks from different view angles for model training. Given the high contrast and resolution of bones in CT, the segmentation masks are of high quality and very time-efficient to be produced, compared to annotating 2D radiographs. (2) All eight carpal bones and two distal forearm bones are included, which will be beneficial for a variety of downstream tasks. (3) The trained segmentation model using over 9000 simulated images and the X-ray simulation code will be freely available to researchers. To the authors' best knowledge, no existing work utilized a similar method to achieve comparable segmentation accuracy for wrist bone segmentation in X-ray images.

## 2. METHOD

### 2.1. CT Dataset and Real X-ray Images

The CT dataset contains 22 subjects (10 female and 12 male, median age 51, age range 25 - 72 years) recruited from the hand clinic at Salford Royal Hospital, Greater Manchester, UK. Each subject was imaged at five different wrist positions: neutral, and four extreme positions in radial-ulnar and flexion extension movement. Hence, a total of 110 CT scans were acquired. After excluding the poor-quality scans, 107 were used in this study. The wrist positions were held on a specially designed foam. Each of the CT volumes is captured by a GE LightSpeed VCT machine with a very low-dose exposure. The acquisition parameters were: tube voltage of 80 kV, focal spot of 0.7 mm, slice thickness of 0.625 mm, pixel spacing of 0.29×0.29 mm$^2$. The volumes were re-sampled by tri-linear interpolation to iso-cubic volumes of 0.5×0.5×0.5 mm$^3$.

The multi-label bone segmentation mask of eight carpal bones and two distal forearm bones (i.e. ulna and radius) was obtained using the interactive image segmentation tool [6] for the neutral pose CT scan of each subject. Then using the shape and kinematic modeling methods proposed by Chen et al. [7], the segmentation masks of all CT scans were obtained.

For method evaluation purposes, 121 real wrist radiographs were collected from Queen's Medical Centre, Nottingham, UK. These images cover different view angles and geometrical variations are available for evaluation. These radiographs do not contain segmentation masks, and they were used for qualitative analysis only.

Subject consensus and ethical approvals for both the CT and X-ray datasets were in place for this study, and all images were anonymized.

### 2.2. X-ray Simulation from CT

The key to the success of using simulated X-ray images to train a model that can be generalized to real images is to ensure the simulated data distribution is as realistic as possible and covers a variety of conditions, including intensity variations, view differences, geometric changes, subject-level differences, and image artifacts. The images of some key steps are shown in Fig. 1.

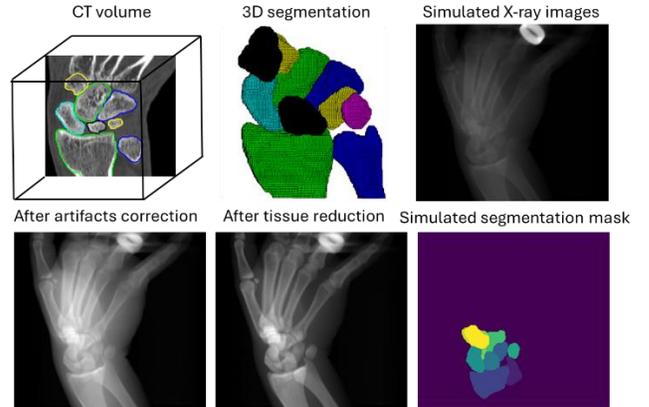

Fig. 1 The key steps of X-ray image simulation.

To capture subject-level differences, our CT data were collected from both males and females, different wrist poses, and a large range of age groups (section 2.1). A statistical wrist model [7] was previously created based on this data, indicating a good representation of the population.

X-ray simulation using CT images is not a new technique, and many works have utilized this process for different applications, such as 3D-2D image registration [8]. In our method, an orthographic projection model is used in X-ray image simulation by assuming all X-rays are parallel to each other. It avoids the complication of simulating different perspective projection settings, and the effects of perspective projection can be approximated using wrist resizing in the data augmentation process described later. It is assumed that the effects of the scatter and differential absorption across the energy spectrum of the X-rays are not significant. The attenuation of an X-ray traversing the human body (or CT volume) can be expressed as [9]:

$$N_{out} = N_{in} e^{-\sum_i u_i d_i} \quad (1)$$

where $N_{in}$ and $N_{out}$ are the number of photons entering and exiting the body. $u_i$ represents the absorption coefficient of the material $i$ and $d$ is the thickness of the material. The image intensity is proportional to $N_{out}/N_{in}$, hence can be estimated by calculating the exponential of the negative sum of CT values along each ray.

Artifact correction is also conducted to deal with unexpected objects that appear in the CT data (e.g. a metal ring in Fig. 1) that could distort the intensity of the simulated intensity. This correction process deals only with extreme situations by setting the maximum value to be the 99$^{th}$ percentile value.

Additionally, when generating X-ray images from different viewpoints, the CT volume is rotated using image interpolation methods. When the volume is rotated by an angle that is not divisible by 90°, some image artifacts may appear if zeros are filled in the empty space of the interpolated CT volume. Instead, the CT values for air should be used to fill in the empty space. In practice, we set all negative intensity values to 0 to offset the negative air CT values.

Tissue reduction is an additional step we proposed to reduce the effects caused by the tissues and to make the bone region more visible. This mimics the process of the post-processing method that X-ray vendors use to enhance bone visibility. In practice, this is implemented by setting the image intensity values that are smaller than the 20th percentile to the 10th percentile, which is determined by trial and error, and assuming each image contains pure tissue regions.

To simulate the geometric changes, X-ray images are generated from different view angles, ranging from -70° to 70° in a 10° interval. This results in 15 unique images per CT volume. 0° represents the AP view of the wrist.

After all the above processing is completed, min-max image normalization is conducted, followed by image resizing (256×256) to train the image segmentation model.

Finally, the corresponding masks of the simulated X-ray images are generated by using the 3D multi-label segmentation masks obtained from the CT scans. The projected label value is determined by the closest class label when bone overlapping occurs, as shown in Fig. 1.

### 2.3. Image Segmentation Model Training

The image segmentation model is trained using the nnUnet method [10]. It is a widely used CNN-based model and has achieved the SOTA performance in many medical image segmentation tasks. It utilizes an encoder-decoder architecture that is self-optimized during model training. nnUnet trains the model with a combination of Dice loss and cross-entropy loss.

Specific to our work, we added a bespoke image augmentation module to nnUnet for further simulating different conditions during the model training. In this case, the image augmentation runs 7 times per input image with randomized settings as follows: rotation (-40° to 40°), translation (maximum 20% of the image's height and width), zoom (maximum +/-20% zoom), and flip (only horizontally). The horizontal flip is one of the most important augmentation processes, as it improves the model's robustness in processing both left and right-hand images.

## 3. EVALUATION

### 3.1. Experimental Design

Among the 22 unique subjects (107 CT scans), 88 CT volumes from 18 subjects were used for model training, and 19 CT scans from the remaining 4 subjects were reserved for testing. In total, 9240 (i.e. 88×15×7) simulated images were used for model training.

The test set was generated using the same process as the training data as described in section 2.2. To evaluate the performance for different view angles, 114 images per angle (ranging from 0° to 70° with an interval of 10°) were generated using the reserved 19 CT scans using the image augmentation process described in section 2.3. A total of 912 (114×8) simulated wrist X-rays were generated for testing.

The Dice score and average surface distance (ASD) measures were used as the evaluation metrics for the simulated test set.

Additionally, the 121 real wrist radiographs were tested for qualitative analysis as no ground truths are available for these real X-ray images due to the challenges in performing manual annotation in difficult view angles, even by experts.

No existing method could segment wrist radiographs in a variety of view angles, hence only our method was evaluated.

### 3.2. Quantitative Evaluation on Simulated X-ray Images

The results of Dice and ASD are reported in Table 1. Note that the results of both negative and positive of the same angle are combined.

**Table 1.** Mean values of Dice and ASD for the simulated test set. All bones across different view angles are reported.

| Bone \ Angle | Dice | | | | | | | |
|---|---|---|---|---|---|---|---|---|
| | 0° | 10° | 20° | 30° | 40° | 50° | 60° | 70° |
| Ulna | 0.92 | 0.92 | 0.92 | 0.90 | 0.88 | 0.87 | 0.82 | 0.74 |
| Radius | 0.95 | 0.94 | 0.93 | 0.93 | 0.92 | 0.91 | 0.89 | 0.86 |
| Triquetrum | 0.86 | 0.87 | 0.89 | 0.90 | 0.88 | 0.87 | 0.84 | 0.76 |
| Lunate | 0.94 | 0.94 | 0.94 | 0.91 | 0.89 | 0.84 | 0.85 | 0.80 |
| Scaphoid | 0.92 | 0.90 | 0.89 | 0.88 | 0.91 | 0.91 | 0.90 | 0.85 |
| Pisiform | 0.93 | 0.94 | 0.95 | 0.94 | 0.93 | 0.90 | 0.89 | 0.85 |
| Hamate | 0.94 | 0.94 | 0.94 | 0.93 | 0.91 | 0.90 | 0.87 | 0.85 |
| Capitate | 0.92 | 0.93 | 0.92 | 0.90 | 0.85 | 0.81 | 0.72 | 0.67 |
| Trapezoid | 0.84 | 0.84 | 0.85 | 0.82 | 0.79 | 0.80 | 0.80 | 0.70 |
| Treapezium | 0.94 | 0.94 | 0.94 | 0.94 | 0.94 | 0.93 | 0.92 | 0.90 |
| Average | 0.92 | 0.92 | 0.92 | 0.90 | 0.89 | 0.87 | 0.85 | 0.80 |
| | ASD | | | | | | | |
| Ulna | 1.24 | 1.56 | 1.27 | 1.47 | 1.55 | 3.39 | 2.56 | 4.33 |
| Radius | 1.70 | 1.70 | 1.61 | 1.65 | 1.74 | 1.98 | 2.45 | 2.57 |
| Triquetrum | 1.14 | 1.11 | 0.82 | 0.84 | 1.04 | 1.09 | 2.39 | 1.98 |
| Lunate | 1.06 | 1.02 | 0.80 | 0.86 | 0.97 | 2.48 | 1.32 | 1.94 |
| Scaphoid | 0.89 | 0.98 | 1.42 | 1.13 | 0.90 | 0.98 | 1.28 | 2.36 |
| Pisiform | 0.84 | 0.65 | 0.62 | 0.67 | 0.81 | 1.20 | 1.12 | 2.15 |
| Hamate | 1.03 | 0.90 | 0.88 | 0.90 | 1.10 | 1.11 | 1.55 | 2.49 |
| Capitate | 1.54 | 1.21 | 1.07 | 1.17 | 1.56 | 1.77 | 2.58 | 3.14 |
| Trapezoid | 1.69 | 1.26 | 1.31 | 1.37 | 1.60 | 1.32 | 1.42 | 2.33 |
| Treapezium | 1.24 | 0.87 | 0.88 | 0.93 | 1.02 | 1.08 | 1.32 | 1.86 |
| Average | 1.24 | 1.13 | 1.07 | 1.10 | 1.23 | 1.64 | 1.80 | 2.52 |

It can be seen that the model performed well when the view angles were smaller than 30° with an average Dice of ~0.92. The performance decreased as the view angle increased, which is expected as the bone interposition became more severe. Moreover, higher performance was achieved on the radius and ulna due to larger bone sizes and not overlapping with the 8 carpal bones. Among the carpal bones, the capitate achieved a high accuracy at smaller view angles with an average Dice of ~0.92. The performance then decreased rapidly as the angle increased, only achieving a Dice score of 0.67. This is expected as the capitate is located at the center of the wrist, so it is more likely to be overlapped with other bones when the view angle increases. The lowest performance was from the Trapezoid, which the model consistently struggles to segment. From qualitative analysis of the segmentations, it seems that the model struggles to define the outline of the Trapezoid in fine detail.

### 3.3. Qualitative Evaluation on Real X-ray Images

The trained model was also applied to the test set that contains 121 real X-ray images. Some visual examples are shown in Fig. 2 which cover different view angles, and different field-of-views for both left and right hands.

It can be seen that the model produced acceptable segmentation results and correct anatomical relationships between the overlapped bones, even at very challenging view angles. The method failed when only part of the hand was visible (Fig.2 red box), as this situation was not simulated in the training process. It can be appreciated that it is extremely challenging for humans to annotate accurate segmentation masks for these real X-ray images to perform quantitative evaluation.

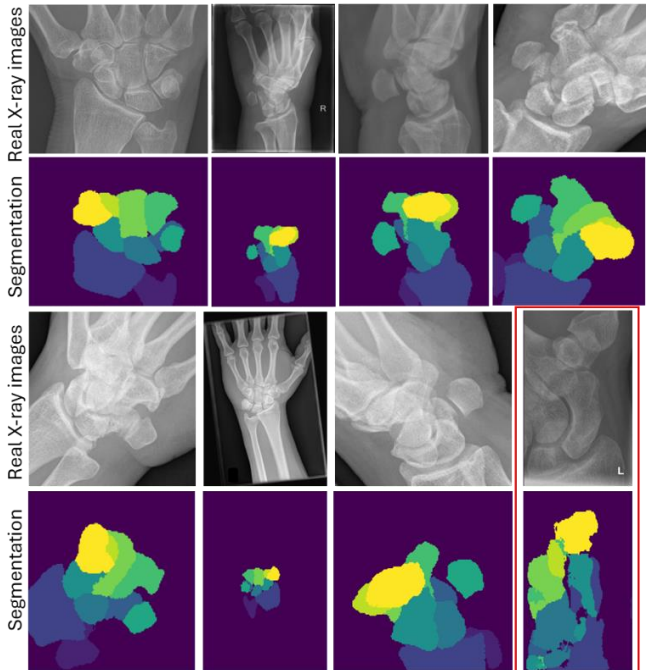

Fig. 2: Segmentation results of real X-ray images of different view angles and field-of-views for left and right hands.

### 4. CONCLUSION

A wrist X-ray image segmentation method has been presented, which can achieve automatic segmentation of ten bones (i.e. 8 carpal bones and 2 distal forearm bones). The method utilized thousands of simulated X-ray images and their corresponding masks generated from CT scans that cover different view angles, field-of-views, gender, and age groups to train a deep learning segmentation model (nnUnet). The method was evaluated quantitatively using simulated images and qualitatively using real X-ray images. The average Dice score of all bones in smaller view angles (close to the AP view) for simulated images is ~0.90 and the qualitative results also demonstrated the superiority of our method. To the author's best knowledge, no existing work has achieved similar performance. We will include more simulated conditions to improve the model performance and produce a dataset to quantitatively evaluate real X-ray images.

### 5. COMPLIANCE WITH ETHICAL STANDARDS

This research study was conducted retrospectively using human subject data made available through local medical sources. Ethical approval was obtained for the use of this data. The subjects used in this study had consented to be included in this research. All data was anonymized, and the participants' information cannot be identified from the imaging data.

### 6. CONFLICTS OF INTEREST

This work was supported by the Medical Research Council, U.K., under Grant 87997.